\documentclass[aps,prd,twocolumn,nofootinbib,superscriptaddress]{revtex4-2}
\usepackage{graphicx}
\usepackage{dcolumn}
\usepackage{bm}
\usepackage{hyperref}
\usepackage{booktabs}
\usepackage{xcolor}
\usepackage[utf8]{inputenc}
\usepackage{soul}
\usepackage{amsmath}
\usepackage{orcidlink}

\usepackage{soul}

\soulregister\ref7

\definecolor{cyan_dp}{HTML}{007272}
\definecolor{rose}{HTML}{a84847}
\definecolor{green}{HTML}{6A8023}
\definecolor{orange}{HTML}{ff5500}
\definecolor{red}{HTML}{FF2511}
\definecolor{blue}{HTML}{055C9D}
\definecolor{orchid}{HTML}{9c3386}
\definecolor{ref_blue}{HTML}{1063b5}
\definecolor{forest_green}{HTML}{2d854c}

\hypersetup{
    colorlinks,
    citecolor=ref_blue,
    filecolor=black,
    linkcolor=forest_green,
    urlcolor=cyan_dp
}

\definecolor{byzantium}{HTML}{bd2ea3}
\newcommand\nmesh{\texttt{Nmesh}}

\begin{document}

\title{A discontinuous Galerkin method with artificial viscosity geared
toward numerical relativity applications}


\author{Wolfgang Tichy\orcidlink{0000-0002-8707-754X}}
\affiliation{Department of Physics, Florida Atlantic University,
                    Boca Raton, FL 33431, USA}

\author{Ananya Adhikari\orcidlink{0000-0002-3890-3577}}
\affiliation{CENTRA, Departamento de F\'isica, Instituto Superior
 T\'ecnico IST, Universidade de Lisboa UL, Avenida Rovisco Pais 1,
 1049 Lisboa, Portugal}

\author{Faezeh Faghihnia Torshizi\orcidlink{0009-0008-4063-5023}}
\affiliation{Department of Physics, Florida Atlantic University,
                    Boca Raton, FL 33431, USA}





%
\newcommand\be{\begin{equation}}
\newcommand\ba{\begin{eqnarray}}

\newcommand\ee{\end{equation}}
\newcommand\ea{\end{eqnarray}}
\newcommand\p{{\partial}}
\newcommand\remove{{{\bf{THIS FIG. OR EQN. COULD BE REMOVED}}}}

%

\begin{abstract}

Artificial viscosity can be used to deal with shocks and other
discontinuities such as star surfaces, that can arise in matter. We explain
in detail how a local discontinuous Galerkin method can be used to add
artificial viscosity terms to hyperbolic evolution equations. These
artificial terms are multiplied by a viscosity that is non-zero only in elements
that are flagged as non-smooth by a newly developed smoothness indicator.
Our artificial viscosity method works well in several simple shock test
cases such as with Burgers' equation and blast wave tests. For the first
time, we are also able to evolve single neutron star in full general
relativistic hydrodynamics. To achieve this, we use different viscosities
for the momentum density than for the rest of the matter
variables.

\end{abstract}

\maketitle

\section{Introduction}

The discontinuous Galerkin (DG) method, which was first introduced by Reed
and Hill~\cite{Reed1973a} and then studied in detail for nonlinear
hyperbolic conservation laws by Cockburn and
Shu~\cite{Cockburn:1991a,Cockburn:1989a,
Cockburn:1989b,Cockburn:1990a,Cockburn1998a}, has proven very useful and
versatile. DG methods have at least two advantages when compared to more
traditional finite volume (FV) or finite difference (FD) methods. They offer
high-order convergence when the fields are smooth, and they reduce the
communication overhead when the computational domain is divided into small
elements that are distributed among different compute cores. Over the last
15 years or so there has been a growing interest to use DG methods in
numerical relativity
problems~\cite{Radice:2011qr,Bugner:2015gqa,Teukolsky:2015ega,
Hilditch:2015aba,
Kidder:2016hev,Dumbser:2017okk,Koppel:2017kiz,Fambri:2018udk,
Cao:2018myc,Deppe:2021ada,
Deppe:2021bhi,Tichy:2022hpa,Deppe:2024ckt,Adhikari:2025nio}.

While DG methods work very well for problems with smooth fields such a black
hole simulations, problems with matter, where discontinuities or shocks can
occur, have proven harder to tackle. Numerous nonlinear
limiters~\cite{Shu:1987a,Cockburn:1989a,Cockburn1998a,Schaal:2015ila,
Krivodonova2007a,Moe:2015a} have been proposed for this purpose. However,
these limiters fall short~\cite{Deppe:2021ada,Deppe:2021bhi,Deppe:2024ckt}
compared to the more robust traditional FV methods, with e.g.~weighted
essentially nonoscillatory (WENO)~\cite{Liu1994a,Jiang1996a} reconstruction,
that are used for high-resolution shock capturing. Therefore, a hybrid
approach~\cite{Costa2007a,Huerta2012a,Sonntag2014a,
Dumbser2014a,Boscheri2017a,Zanotti:2015mia,Fambri:2018udk,
NunezdelaRosa2018a,Deppe:2021ada,Deppe:2024ckt,Adhikari:2025nio} is often
taken where DG is used in all regions with smooth fields and FV is only
used in non-smooth regions, which are detected according to certain
``trouble'' criteria.
While such hybrid DG/FV methods have recently been used successfully
for single~\cite{Deppe:2021ada,Deppe:2021bhi,Adhikari:2025nio} and even
binary neutron stars~\cite{Deppe:2024ckt}, they introduce additional
complications and overheads since information has to be communicated between
elements that use different methods (DG vs.~FV) and different grid points.
For this reason we consider an alternative treatment here, namely
artificial viscosity (AV).
This approach, first introduced in the context of FD methods by Richtmyer
and Von Neumann~\cite{Margolin2024a,VonNeumann1950a}, adds extra viscosity
terms to the equations of hydrodynamics to simplify the treatment of shocks.
The AV approach for DG methods, that has gained interest in the
computational physics and applied mathematics literature (see
e.g.~\cite{Persson2006a,Kloeckner2011a,Barter2010a,Zingan2013a,Yu2020a}), is
largely based on~\cite{Tadmor1989a}. In its modern form the extra AV terms
are multiplied by a viscosity $\mu$, which is chosen to be non-zero only in
elements that contain discontinuities in the evolved fields.

DG methods with AV are well established by now. Nevertheless, they have not
been extensively used in numerical relativity yet. For example,
while~\cite{Anninos_2017} has conducted various tests for hydrodynamics using
a DG method with AV, no attempt was made to evolve self-gravitating systems
such as neutron stars. The more recent work in~\cite{Atteneder2025a} has also
conducted many successful tests, but has ultimately failed in its attempt to
stably evolve single neutron stars.

The main purpose of this paper is to describe and test a new artificial
viscosity (AV) approach in the context of numerical relativity. It is based
on a local discontinuous Galerkin (LDG) method~\cite{Cockburn1998b} that we
have implemented in our
\nmesh~\cite{Tichy:2022hpa,Adhikari:2025nio,Adhikari2026a} code.
The LDG method can be used to deal with evolution equations that contain
second spatial derivatives, which arise from the extra AV terms. As in other
works the AV terms are multiplied by a viscosity $\mu$, which is non-zero only
in elements with non-smooth fields. The factor $\mu$ is computed from a new
smoothness indicator $\tau$ whose construction we describe in detail. Our new
AV prescription works in various test cases, and, for the first time, we can
successfully evolve a neutron star in full general relativistic
hydrodynamics (GRHD) using AV within a DG method.

In Sec.~\ref{num_methods} we describe the LDG method and other numerical
methods we use.
In Sec.~\ref{LDG_tests} we test the LDG method for a simple scalar wave.
Section~\ref{ArtViscosity} shows how the LDG method can be used to add
artificial viscosity (AV) terms to evolution equations.
Next, in Sec.~\ref{av_tests} we consider Burgers' equations and the
equations of general relativistic hydrodynamics to test our new AV approach.
We summarize and discuss our results in Sec.~\ref{discussion}.
Throughout the article, we use geometric units, in which $G=c=1$, as well as
$M_{\odot}=1$.

\section{Numerical methods}
\label{num_methods}

In this section, we present the numerical method we use to perform
simulations with hyperbolic or parabolic evolution equations.

\subsection{The discontinuous Galerkin method}

As explained in~\cite{Tichy:2022hpa,Adhikari:2025nio}, discontinuous
Galerkin (DG) methods are often used for systems of evolution equations that
are first order in space, no matter if they are in conservative form or not.
In the conservative case the equations take the form
\be
\label{Cons1}
\partial_t u +  \partial_i g^i = s ,
\ee
where $u$ denotes the conserved matter variables with corresponding fluxes
$g^i = g^i(u)$ and sources $s=s(u)$.
However, many equations of interest cannot be brought in this form at all.
Consider for example the heat equation
\be
\label{Heat1}
\partial_t u =  \partial_i \partial_i u .
\ee
If we define a new variable $q^i = \partial_i u$, Eq.~(\ref{Heat1}) is
equivalent to
\ba
\label{Heat2_1}
\partial_t u -  \partial_i q^i  &=&  0, \\
\label{Heat2_2}
q^i &=& \partial_i u .
\ea
Thus the first equation seems to be of the same form as in
Eq.~(\ref{Cons1}). Note, however, that $q^i$ will depend on the derivative
of $u$, unlike the $g^i(u)$ above. More importantly, if we were to
discretize Eq.~(\ref{Heat2_1}) using the DG scheme appropriate for
Eq.~(\ref{Cons1}), while computing $q^i$ using a differentiation matrix
(as described in~\cite{Tichy:2022hpa}), we obtain an unstable scheme.
It is therefore clear that we need a different discretization scheme.
Let us look at the more general case
\ba
\label{Gen_1}
\partial_t u + \partial_i g^i - \partial_i (\mu q^i)  &=&  s, \\
\label{Gen_2}
q^i &=& \partial_i u ,
\ea
where $g^i = g^i(u)$, $s=s(u)$, $\mu$ is an adjustable function, and
$q^i$ depends on the derivative of $u$.

To discretize the spatial derivatives in Eq.~(\ref{Gen_1}), we first
integrate against a test function $\psi$ over the volume $\Omega$ of
one element.
We obtain
\be
\label{Integ0}
\int (\psi\partial_t u + \psi\partial_i f^i) d^3x = \int \psi s d^3x ,
\ee
where we use the abbreviation
\be
f^i := g^i - \mu q^i .
\ee
The second term is now integrated by parts using
\be
\label{int_by_parts}
\int \psi\partial_i f^i d^3x =
\oint \psi f^i n_i d^2\Sigma - \int f^i \partial_i\psi d^3x ,
\ee
where $d^2\Sigma$ is the surface element for integrating over the
boundary $\partial\Omega$, and $n_i$ is normal to $\partial\Omega$.
To incorporate numerical boundary conditions, the flux $f^i n_i$
in the surface integral over the boundary is replaced by the
numerical flux $(f^i n_i)^*$ (See Sec.~\ref{numflux} below).
The replacement of $f^i n_i$ by $(f^i n_i)^*$ in the surface integral yields
\be
\label{weak_form}
\int \psi\partial_i f^i d^3x
\rightarrow
\oint \psi (f^i n_i)^* d^2\Sigma - \int f^i \partial_i\psi d^3x .
\ee
This leads to the weak from of the equations. We integrate
by parts again to eliminate derivatives of $\psi$, and arrive at
\be
\label{strong_form}
\int \psi\partial_i f^i d^3x
\rightarrow
\oint \psi [(f^i n_i)^* - f^i n_i] d^2\Sigma + \int \psi\partial_i f^i d^3x .
\ee
The latter is known as the strong form, which is what we will use in this work.
With this replacement, Eq.~(\ref{Integ0}) becomes
\be
\label{Integ_1}
\int (\psi\partial_t u + \psi\partial_i f^i) d^3x
= \int \psi s d^3x - \oint \psi [(f^i n_i)^* - f^i n_i] d^2\Sigma .
\ee

So far the discussion has followed the same steps that one would take
to construct an ordinary DG method. In order to arrive at a
local discontinuous Galerkin (LDG) method~\cite{Cockburn1998b} we
follow the ideas in~\cite{Hesthaven2008} and~\cite{Zhang2003a}.
We first rewrite Eq.~(\ref{Gen_2}) as
\be
\label{Gen_2f}
q^k + \partial_i f^{ik} = 0 ,
\ee
where
\be
f^{ik} := -\delta^{ik} u .
\ee
If we now retrace the above steps using Eq.~(\ref{Gen_2f}), we find
\be
\label{Integ_2}
\int (\psi q^k + \psi\partial_i f^{ik}) d^3x
= - \oint \psi [(f^{ik} n_i)^* - f^{ik} n_i] d^2\Sigma .
\ee
We see that both Eq.~(\ref{Integ_1}) and (\ref{Integ_2}) have almost the
same form.

To arrive at the final spatial discretization of Eq.~(\ref{Integ_1}) we use
Lagrange’s characteristic polynomials for the test function $\psi$ and also
expand all fields in terms of these polynomials.
Following all the steps that lead from Eq.~(9) to Eq.~(28)
in~\cite{Tichy:2022hpa} we find
\ba
\label{Discr_1}
    \partial_t u_{q_1 q_2 q_3}
  + \sum_{r=0}^N \Big(
    \frac{\partial x^{\bar{1}}}{\partial x^i}
    D_{q_1 r}^{\bar{1}} f^i_{r q_2 q_3}
  + \frac{\partial x^{\bar{2}}}{\partial x^i}
    D_{q_2 r}^{\bar{2}} f^i_{q_1 r q_3}                 \ \ && \nonumber\\
  + \frac{\partial x^{\bar{3}}}{\partial x^i}
    D_{q_3 r}^{\bar{3}} f^i_{q_1 q_2 r}
  \Big)                                                     && \nonumber\\
= s_{q_1 q_2 q_3}
 -\frac{\sqrt{\bar{\gamma}^{\bar{1}\bar{1}}_{q_1 q_2 q_3}}}{w_{q_1}}
  F_{q_1 q_2 q_3} (\delta_{q_1 0} + \delta_{q_1 N})   \ \ \ && \nonumber\\
 -\frac{\sqrt{\bar{\gamma}^{\bar{2}\bar{2}}_{q_1 q_2 q_3}}}{w_{q_2}}
  F_{q_1 q_2 q_3} (\delta_{q_2 0} + \delta_{q_2 N})   \ \ \ && \nonumber\\
 -\frac{\sqrt{\bar{\gamma}^{\bar{3}\bar{3}}_{q_1 q_2 q_3}}}{w_{q_3}}
  F_{q_1 q_2 q_3} (\delta_{q_3 0} + \delta_{q_3 N}) , \ \ \
\ea
where
\be
\label{def_F}
F := (f^i n_i)^* - f^i n_i ,
\ee
and $(f^i n_i)^*$ denotes the numerical flux, computed from the
flux in the local element and the adjacent element. To arrive at
Eq.~(\ref{Discr_1}) we have performed Gau{\ss}ian quadrature using
$N+1$ Gau\ss-Lobatto grid points in each of the three directions.
The grid point coordinates in each local element are denoted by
$x^{\bar{i}}_{q_1 q_2 q_3}$ where $q_i \in [0,N]$ is the grid point index
in each direction $i$. Here $u_{q_1 q_2 q_3}$ and $f^i_{q_1 q_2 q_3}$
denote $u$ and $f^i$ at grid point $x^{\bar{i}}_{q_1 q_2 q_3}$,
and $w_q$ are the Gau\ss~quadrature weights.
For $q \neq r$ the differentiation matrix $D_{qr}^{\bar{x}}$ is given by
\be
D_{qr}^{\bar{x}} = \frac{c_r}{c_q}\frac{1}{\bar{x}_q - \bar{x}_r} ,
\ee
in terms of the Lagrange interpolation weights
\be
c_q = \left[\prod_{s=0, s\neq q}^N(\bar{x}_q - \bar{x}_s)\right]^{-1} ,
\ee
and $D_{qq}^{\bar{x}}$ can be obtained from $\sum_r D_{qr}^{\bar{x}} = 0$.
Note that we have also performed a coordinate transformation
\be
\label{x_of_Xbar}
x^{\bar{i}} = x^{\bar{i}}(x^i)
\ee
from Cartesian coordinates $x^i$ to local element coordinates $x^{\bar{i}}$,
such that the element ranges from $-1$ to $+1$ in all three
$x^{\bar{i}}$ coordinates.

Similarly, Eq.~(\ref{Integ_2}) becomes
\ba
\label{Discr_2}
q^k_{q_1 q_2 q_3}
&=&
    \sum_{r=0}^N \Big(
    \frac{\partial x^{\bar{1}}}{\partial x^k}
    D_{q_1 r}^{\bar{1}} u_{r q_2 q_3}
  + \frac{\partial x^{\bar{2}}}{\partial x^k}
    D_{q_2 r}^{\bar{2}} u_{q_1 r q_3}                  \nonumber\\
&&
  + \frac{\partial x^{\bar{3}}}{\partial x^k}
    D_{q_3 r}^{\bar{3}} u_{q_1 q_2 r}
  \Big)                                                \nonumber\\
&&
 -\frac{\sqrt{\bar{\gamma}^{\bar{1}\bar{1}}_{q_1 q_2 q_3}}}{w_{q_1}}
  F^k_{q_1 q_2 q_3} (\delta_{q_1 0} + \delta_{q_1 N})  \nonumber\\
&&
 -\frac{\sqrt{\bar{\gamma}^{\bar{2}\bar{2}}_{q_1 q_2 q_3}}}{w_{q_2}}
  F^k_{q_1 q_2 q_3} (\delta_{q_2 0} + \delta_{q_2 N})  \nonumber\\
&&
 -\frac{\sqrt{\bar{\gamma}^{\bar{3}\bar{3}}_{q_1 q_2 q_3}}}{w_{q_3}}
  F^k_{q_1 q_2 q_3} (\delta_{q_3 0} + \delta_{q_3 N}) ,
\ea
with
\be
\label{def_Fk}
F^k := (f^{ik} n_i)^* - f^{ik} n_i = -(u n_k)^* + u n_k .
\ee

As we can see Eq.~(\ref{Discr_2}) does not contain only the differentiation
matrices $D_{q_1 r}^{\bar{i}}$ that correspond to the derivative terms in
Eq.~(\ref{Gen_2}), but also extra terms on the right-hand side. The
latter terms are essential for numerical stability.

\subsection{Numerical fluxes for the local discontinuous Galerkin method}
\label{numflux}

The discretizations in Eqs.~(\ref{Discr_1})~and~(\ref{Discr_2})
contain the numerical fluxes
\be
(f^i n_i)^* = (g^i n_i)^* - (\mu q^i n_i)^*
\ee
and
\be
(f^{ik} n_i)^* = -(u n_k)^*
\ee
that we have not specified yet. For $(g^i n_i)^*$, it is natural to simply
use the same flux that we would use for the standard $\mu=0$ case
without second derivatives, i.e., the local Lax-Friedrichs (LLF) flux
\be
\label{LLF_flux}
(g^i n_i)^* = \frac{1}{2}\left[
  g^i(u_\mathrm{in}) n_i + g^i(u_\mathrm{adj}) n_i
  + |\lambda|_\mathrm{max}\left(u_\mathrm{in} - u_\mathrm{adj}\right) \right] ,
\ee
or possibly the upwind flux.
Here $u_\mathrm{in}$ is the field value at the boundary using grid points
that belong to the element enclosed by the boundary, $u_\mathrm{adj}$ is the
field value at the boundary using grid points that belong to the adjacent
element on the other side of the boundary, $n_i$ is the outward pointing
normal to the boundary, and $|\lambda|_\mathrm{max}$ is the absolute value
of the eigenvalue of the characteristic mode with the largest eigenvalue
magnitude, considering eigenvalues from both sides.

For the other two terms, we follow~\cite{Hesthaven2008} and use central
fluxes given by
\be
\label{central_flux1}
(u n_k)^* = \frac{1}{2}\left[
             u_\mathrm{in} n_k + u_\mathrm{adj} n_k \right]
\ee
and
\be
\label{central_flux2}
(\mu q^i n_i)^* = \frac{1}{2}\left[
  (\mu q^i)_\mathrm{i} n_i + (\mu q^i)_\mathrm{a} n_i \right] .
\ee
Here $(\mu q^i)_\mathrm{i}$ is the $\mu q^i$ constructed in the element
and $(\mu q^i)_\mathrm{a}$ is the $\mu q^i$ constructed in the adjacent
element.
Note that $(\mu q^i)_\mathrm{i}$ and $(\mu q^i)_\mathrm{a}$ need to be
computed using Eq.~(\ref{Discr_2}), i.e., from the fields $u_\mathrm{in}$
and $u_\mathrm{adj}$ on either side of the boundary multiplied by
differentiation matrices, and also from $(u n_k)^*$ (contained in
the $F^k$) that also depends on $u_\mathrm{in}$ and
$u_\mathrm{adj}$ from either side.

The discretizations in Eqs.~(\ref{Discr_1})~and~(\ref{Discr_2}) together
with the numerical fluxes in Eqs.~(\ref{LLF_flux}), (\ref{central_flux1})
and (\ref{central_flux2}) comprise the LDG method. The term local in LDG
refers to the fact that, while Eqs.~(\ref{Discr_1})~and~(\ref{Discr_2}) are
coupled, $(u n_k)^*$ is independent of $q^k$. Thus, we can first find $q^k$
from Eqs.~(\ref{Discr_2}) and (\ref{central_flux1}) locally, from
$u_\mathrm{in}$ in an element and $u_\mathrm{adj}$ in the nearest neighbors
of the element. Next we use this $q^k$ in Eq.~(\ref{Discr_1}) to update $u$
in each element. Thus, Eqs.~(\ref{Discr_1}) and ~(\ref{Discr_2}) do not
have to be solved simultaneously and globally.

Nevertheless, the LDG method is more complicated to implement than an
ordinary DG method. For a plain DG method, we only need the field values
$u_\mathrm{in}$ in each element and $u_\mathrm{adj}$ from neighboring
elements to update the evolved field $u$. Therefore, when we parallelize
(e.g. with MPI), we only need to send $u_\mathrm{adj}$ from all neighbors to
each element. For LDG, however, the situation is more complicated. We first
send $u_\mathrm{adj}$ from all neighbors to each element, so that we have
the information to construct $q^k$ from Eqs.~(\ref{Discr_2}) and
(\ref{central_flux1}). Next we need to send $(\mu q^i)_\mathrm{a}$ from all
neighbors to each element. Only after this can we update the evolved field $u$
using Eqs.~(\ref{Discr_1}) and (\ref{central_flux2}). This means more
communication is needed. The only exception to this is if we choose a
$\mu(\bar{x})$ that is always zero on all element boundaries so that $(\mu
q^i n_i)^*=0$, as in e.g.~in the Gegenbauer or super Gaussian viscosity
cases discussed below.


\subsection{Evolution forward in time}
\label{time_evo}

Equations~(\ref{Discr_1}) and (\ref{Discr_2}) still contain time
derivatives and thus form a set of coupled
ordinary differential equations (ODEs) for the fields $u_{q_1 q_2 q_3}$ at
the grid points.
In this work we use standard Runge-Kutta time integrators
to find the solution of these ODEs from the $u_{q_1 q_2 q_3}$ at the initial
time. Such Runge-Kutta methods are only stable when the time step
$\Delta t$ is small enough for
Courant-Friedrichs-Lewy (CFL) condition to be satisfied.
For hyperbolic equations this means
\be
\label{cfl_hyp}
\Delta t  \leq  \Delta x_\mathrm{min} / v ,
\ee
where $\Delta x_\mathrm{min}$ is the distance between the two closest grid
points and $v$ is a number that needs to be greater than the largest
characteristic speed.

Notice however, that Eqs.~(\ref{Gen_1}) and (\ref{Gen_2}) are of parabolic
form for non-zero $\mu$. Then the condition for stability becomes
\be
\label{cfl_para}
\Delta t  \leq  (\Delta x_\mathrm{min})^2 / \mu ,
\ee
which can be much more restrictive.

Even when the CFL condition is satisfied, instabilities can still occur in
some cases. We cure these instabilities by applying the same exponential
filters as in ~\cite{Tichy:2022hpa} and ~\cite{Adhikari:2025nio}.

\section{Tests with the LDG method}
\label{LDG_tests}

In order to verify that the LDG method is implemented correctly in
\nmesh\ we have tested it with both hyperbolic and parabolic equations.

\subsection{The scalar field equation}
\label{Scalar_tests}

We consider a single scalar field obeying the wave equation
\be
\partial_t^2\phi = \partial_i\partial_i\phi .
\ee
In order to apply the LDG method described above we
introduce the extra variable
\be
\Pi := \partial_t\phi .
\ee
This results in the following system equations
\ba
\label{scalar_evo}
\label{phi_evo}
\partial_t\phi                    &=&  \Pi,  \\
\label{Pi_evo}
\partial_t\Pi - \partial_i q^i    &=&  0,    \\
\label{q_phi}
q^k                               &=& \partial_k \phi ,
\ea
where we have again introduced an extra variable $q^k$.
Equation~(\ref{Pi_evo}) is of the form of Eq.~(\ref{Gen_1}) with $g^i=0$ and
$\mu=1$, and Eq.~(\ref{q_phi}) is very similar to Eq.~(\ref{Gen_2}), except
that the right-hand side consists of derivatives of $\phi$ and not $\Pi$.
In addition, we now also have Eq.~(\ref{phi_evo}).
If we define the fluxes
\ba
f_{\phi}^j   &:=& 0.                \nonumber \\
f_{\Pi}^i    &:=& -q^i,             \nonumber \\
f^{ik}       &:=& -\phi\delta^{ik},
\ea
we can use Eq.~(\ref{Discr_1}) together with the LLF flux in
Eq.~(\ref{LLF_flux}) for $g^i=0$ and the central flux in
Eq.~(\ref{central_flux2}) to discretize Eqs.~(\ref{phi_evo})
and~(\ref{Pi_evo}), and Eq.~(\ref{Discr_2}) and the central flux
in Eq.~(\ref{central_flux1}) to discretize Eq.~(\ref{q_phi}).

We have implemented the scalar field equation in three dimensions
following the above recipe.
We find stable evolutions provided we use the same filter
as in~\cite{Tichy:2022hpa} and choose the time step
according to Eq.~(\ref{cfl_hyp}) with $v=8$.
\begin{figure}
\centering
\includegraphics[width=0.99\linewidth]{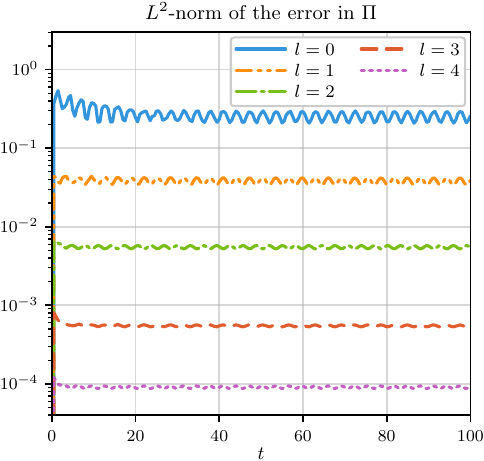}
\caption{
\label{ScalWavPiErr}
The error in $\Pi$ versus time, for a varying number $l$ of
h-refinement levels.
}
\end{figure}
For these evolutions we have used the same filter, domain, initial data, and
boundary conditions as in Sec.~4.1 of~\cite{Tichy:2022hpa}.
Fig.~\ref{ScalWavPiErr} shows the $L^2$-norm of the error in $\Pi$, where we
use $5\times5\times5$ points in each element for a varying number $l$ of
h-refinement levels. We see that the error converges to 0 as expected when
we increase $l$.

\subsection{The heat equation}
\label{Heat_eqn_tests}

When we evolve the 3-dimensional heat equation in the form given in
Eqs.~(\ref{Heat2_1}) and (\ref{Heat2_2}) using the LDG scheme explained
in Sec.\ref{num_methods}, we also find stable and convergent numerical
solutions provided we choose the time step according to
Eq.~(\ref{cfl_para}), with $\mu=1$.


\section{Artificial viscosity}
\label{ArtViscosity}

We have seen that the LDG method for the strong form of the equations,
that is implemented in \nmesh, works as expected for both dissipative
systems like the heat equation and also for hyperbolic systems like the
scalar wave equation.

Our next goal is to add artificial viscosity (AV) to hyperbolic evolution
equations in order to deal with shocks.

When we add viscosity to equations of the form in Eq.~(\ref{Cons1}),
we find
\be
\label{evo_with_art_visc}
\partial_t u +  \partial_i g^i  = s + \partial_i (\mu\partial_i u),
\ee
i.e.~the equations directly fit into the form given in
Eqs.~(\ref{Gen_1}) and (\ref{Gen_2}), if we again introduce $q^i$.

\subsection{Choosing the amount of artificial viscosity}
\label{set_mu}

The main question now is how the value of $\mu$ in
Eq.~(\ref{evo_with_art_visc}) should be chosen. After all we want to
simulate equations that do not have viscosity by themselves. This means
$\mu$ should be non-zero only in regions with shocks, and even there it
should be small enough so that the dissipative piece has no strong influence
on the physics. This means at higher resolutions $\mu$ should converge to
zero.
Following~\cite{Yu2020a,Atteneder2025a} we incorporate this by choosing
$\mu \leq \mu_{\mathrm{max}}$, where
\be
\mu_{\mathrm{max}} = c_{\mathrm{max}} h/(n-1).
\ee
Here $c_{\mathrm{max}}$ is a constant we choose to be of order one,
$h$ is the size of an element, and $n$ is the number of grid points
in one direction. Thus, $\mu_{\mathrm{max}}$ is first order convergent
to zero if we decrease $h$ or increase $n$.

If we were to set $\mu = \mu_{\mathrm{max}}$ everywhere, dissipation
would occur everywhere, which is not needed in smooth regions.
We therefore need a smoothness indicator that can be used to regulate the
amount of viscosity $\mu$. As in~\cite{Yu2020a,Atteneder2025a} we base
this indicator on an estimate of the falloff rate of the
coefficients $c_{l_1 l_2 l_3}$ in the expansion
\be
u(\bar{x})
= \sum_{l_1=0}^N\sum_{l_2=0}^N\sum_{l_3=0}^N c_{l_1 l_2 l_3}
   P_{l_1}(x^{\bar{1}}) P_{l_2}(x^{\bar{2}}) P_{l_3}(x^{\bar{3}})
\ee
in each element. In one dimension, if $u$ is smooth, we expect
the coefficients to approximately fall off like
\be
\label{assumed_falloff}
|c_{l}| = k l^{-\tau},
\ee
where $k$ is a constant. The smoother $u$ is,
the higher the exponent $\tau$ will be.
We therefore choose $\mu$ to be given by
\be
\label{mu_from_tau}
\mu = \mu_{\mathrm{max}} \nu(\bar{x})
\begin{cases}
1					& \text{if } \tau < n_L \\
\left(1-\frac{\tau-n_L}{n_H-n_L}\right)	& \text{if } n_L \leq \tau \leq n_H \\
0					& \text{otherwise},
\end{cases}
\ee
where $n_L$ and $n_H$ are two numbers that we typically choose to be about
1 and 3, and $0\leq \nu(\bar{x})\leq 1$ is a function that we can choose
to give $\mu$ a particular spatial dependence in each element.

The final step is to find an estimate for $\tau$.
We first calculate the Persson
indicator~\cite{Persson2006a} from the sums
\be
\label{c2_sum}
c^2_\mathrm{sum} =
  \sum_{l_1 = 0}^{\bar{N}}
  \sum_{l_2 = 0}^{\bar{N}}
  \sum_{l_3 = 0}^{\bar{N}}c_{l_1 l_2 l_3}^2 ,
\ee
and
\be
\label{c2_hi}
c^2_{\bar{N}} =
   \sum_{l_1 = 0}^{\bar{N}}
     \sum_{l_2 = 0}^{\bar{N}} c_{l_1 l_2 \bar{N}}^{2}
  +\sum_{l_1 = 0}^{\bar{N}}
     \sum_{l_3 = 0}^{\bar{N}} c_{l_1 \bar{N} l_3}^{ 2}
  +\sum_{l_2 = 0}^{\bar{N}}
     \sum_{l_3 = 0}^{\bar{N}} c_{\bar{N} l_2 l_3}^{ 2},
\ee
where $\bar{N}$ is the number of coefficients that remain unfiltered.
The Persson indicator is then
\be
s_{\bar{N}} = \log_{10}( c^2_{\bar{N}} / c^2_\mathrm{sum} ) .
\ee
We see that the faster the $c_{l_1 l_2 l_3}$ fall off, the lower
$s_{\bar{N}}$ will be.

Consider the one dimensional case with coefficients falling off according to
Eq.~(\ref{assumed_falloff}). In this case
$c^2_\mathrm{sum} = c_0^2 + k^2 \sum_{l=1}^{\bar{N}} l^{-2\tau}$.
If we set $c_0=0$ (i.e.~subtract the average), and take into account that
the p-series $\sum_{l=1}^{\infty} l^{-2\tau} \approx 1$ for
$\tau \gtrsim 2$, we find $c^2_\mathrm{sum} \approx  k^2$.
At the same time $c^2_{\bar{N}} = k^2 \bar{N}^{-2\tau}$, thus we obtain
\be
s_{\bar{N}} \approx \log_{10}(\bar{N}^{-2\tau})
= -2\tau\log_{10}(\bar{N}) .
\ee
We will therefore simply define the $\tau$ that we use in
Eq.~(\ref{mu_from_tau}) as
\be
\label{tau_from_Persson}
\tau_{\bar{N}} = -\frac{s_{\bar{N}}}{2\log_{10}(\bar{N})} .
\ee
However, in some special cases with symmetries it can happen
that $c^2_{\bar{N}}$ is almost zero. In that case $s_{\bar{N}}$
does not accurately reflect the true falloff, and $\tau_{\bar{N}}$
is too large.
For this reason we construct a second indicator
\be
s_{\bar{\bar{N}}} = \log_{10}( c^2_{\bar{\bar{N}}} / c^2_\mathrm{sum} ) ,
\ee
where $c^2_{\bar{\bar{N}}}$ is computed from Eq.~(\ref{c2_hi})
with $\bar{N}$ replaced by $\bar{\bar{N}} = \bar{N}-1$. Using
Eq.~(\ref{tau_from_Persson}) with $\bar{N}$ replaced by $\bar{\bar{N}}$
we can get a second $\tau$ given by
\be
\label{tau2_from_Persson}
\tau_{\bar{\bar{N}}} = -\frac{s_{\bar{\bar{N}}}}{2\log_{10}(\bar{\bar{N}})} .
\ee
Our final procedure is then to use the smoothness indicator
\be
\label{min_tau}
\tau = \min(\tau_{\bar{N}}, \tau_{\bar{\bar{N}}})
\ee
in Eq.~(\ref{mu_from_tau}).

In summary, this means the artificial viscosity $\mu$ will be completely
turned off if the estimated falloff $\tau > n_H$, and it will smoothly rise
to $\mu_{\mathrm{max}}$ as $\tau$ drops.
In~\cite{Persson2006a} $\mu$ is chosen to be constant in each element.
However, in~\cite{Glaubitz2019a} it is argued that such a piecewise constant
$\mu$ violates the principle of conservation. Several choices for
the spatial dependence $\nu(\bar{x})$ of $\mu$ in Eq.~(\ref{mu_from_tau})
are discussed in~\cite{Glaubitz2019a}, that are zero at the element
boundaries, and close to one inside the element. In one dimension,
two examples are the Gegenbauer viscosity
\be
\label{GegenbauerVisc}
\nu_{1, \mathrm{Gegenbauer}}(\bar{x}) = (1 - \bar{x}^2)^{\lambda}
\ee
with $\lambda\leq 1$, and the super Gaussian viscosity
\be
\label{SuperGaussianVisc}
\nu_{1, \mathrm{super Gaussian}}(\bar{x}) = e^{-36 \bar{x}^{2\lambda}} ,
\ee
with $\lambda\geq 10$.
In both cases, $\lambda$ can be adjusted to get a $\nu$ that is very
close to one inside ($|\bar{x}|<1$) the element.
In three dimensions we use
\be
\nu = \nu_1(x^{\bar{1}}) \nu_1(x^{\bar{2}}) \nu_1(x^{\bar{3}}) ,
\ee
where $\nu_1$ is whatever one dimensional viscosity we choose.

\section{Testing our artificial viscosity prescription}
\label{av_tests}

\subsection{Burgers' equation}

As a first test we consider Burgers' equation
\be
\partial_t u +  \partial_x u^2/2  = \partial_x (\mu \partial_x u)
\ee
in one dimension, where we have already added the AV
term on the right-hand side.
As in~\cite{Leveque92,Atteneder2025a} we choose initial data of the form
\be
u(x,t=0) = \sin(2\pi x)
\ee
and evolve it on a domain $x \in [0,1]$. As long as $t<1/(2\pi)$,
a smooth analytic solution for $u(x,t)$ can be found by solving
\be
u = \sin(2\pi(x - t u))
\ee
for $u$. At $t=1/(2\pi)$~\cite{Leveque92} a shock front forms at $x=0.5$.
Thus, $\mu$ should become non-zero near $x=0.5$ around this time, to
avoid Gibbs oscillations.

\begin{figure}
\centering
\includegraphics[width=0.99\linewidth]{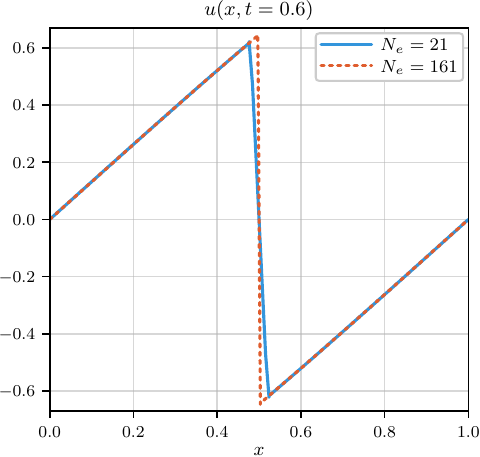}
\caption{
\label{Burgers1_u_t0.6}
The field $u$ at $t=0.6$ for different numbers of elements $N_{\mathrm{e}}$,
with 5 grid points per element.
}
\end{figure}
We numerically evolve this initial data until $t=0.6$ for a varying number
of elements $N_{\mathrm{e}}$ with 5 grid points per element.
In Fig.~\ref{Burgers1_u_t0.6} we see that with increasing resolution
the shock is resolved better. For clarity, we only show the lowest and
highest resolution.
\begin{figure}
\centering
\includegraphics[width=0.99\linewidth]{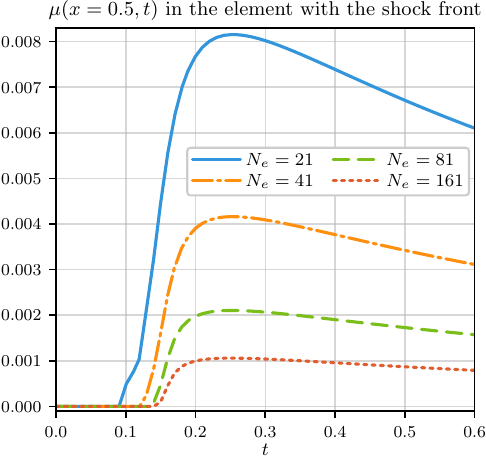}
\caption{
\label{Burgers1_mu_x0.5}
The value of $\mu$ at $x=0.5$ in the element containing the shock front
for different numbers of elements $N_{\mathrm{e}}$ with
$c_{\mathrm{max}}=1$, $n_L=1$, $n_H=3$, and the Gegenbauer
viscosity with $\lambda=1$.
}
\end{figure}
In this test case the shock is not moving and stays at $x=0.5$. This is
correctly detected by the smoothness indicators discussed in
Sec.~\ref{set_mu}. For the viscosity shapes in Eqs.~(\ref{GegenbauerVisc})
and (\ref{SuperGaussianVisc}), $\mu$ is zero everywhere at all times,
except in the central element that contains $x=0.5$.
Fig.~\ref{Burgers1_mu_x0.5} shows the value of $\mu$ in the center
of this shocked element for the Gegenbauer viscosity with $\lambda=1$.
We see that
$\mu$ rises just before $t=1/(2\pi)$ when the shock forms. As designed,
$\mu$ converges to zero at first order with increasing resolution
$N_{\mathrm{e}}$.

On the other hand, if we choose a constant viscosity shape
($\nu(\bar{x})=1$) in each element, we find that $\mu$ develops a small but
non-zero value also in the elements directly left and right of the shock.

\subsection{General relativistic hydrodynamics}

The equations of general relativistic hydrodynamics (GRHD)
can be brought into the flux conservative form of Eq.~(\ref{Cons1}).
Here we use the Valencia formulation in the exact form as
in~\cite{Adhikari:2025nio}, that is given in terms of the
rest-mass density $D$, the internal energy density $\tau$, and the momentum
density $S_j$ as seen by Eulerian observers.
When we add AV terms, their evolution equations become
\ba
\label{GRHD_with_av}
\partial_t D + \partial_i g^i_D       &=& \partial_i(\mu\partial_i D),
\nonumber \\
\partial_t\tau + \partial_i g^i_{\tau}&=& s_{\tau}+\partial_i(\mu\partial_i\tau),
\nonumber \\
\partial_t S_j + \partial_i g^i_{S_j} &=& s_{S_j}+\partial_i(\mu_S\partial_i S_j)
,
\ea
where the fluxes $g^i_D$, $g^i_{\tau}$, $g^i_{S_j}$ and the sources
$s_{\tau}$, $s_{S_j}$, given e.g.~in~\cite{Tichy:2022hpa,Adhikari:2025nio},
depend on the rest-mass energy density $\rho_0$, the specific internal
energy $\epsilon$, the velocity $v^i$, and the pressure $P$.
Together $(\rho_0, \epsilon, v^i, P)$ are known as the primitive variables.
Note that here we have introduced $\mu$ and $\mu_S$ so that we can use
a different amount of artificial viscosity for $S_j$.
Finally, Eq.~(\ref{GRHD_with_av}) has to be brought into the form
of Eqs.~(\ref{Gen_1}) and (\ref{Gen_2}) by introducing $q^i$-variables.

\subsubsection{Blast wave tests}

As a first test we conduct a special relativistic blast wave test in one
dimension. In this case the metric variables are held constant at their
Minkowski space values, so that the sources $s_{\tau}$ and $s_{S_j}$ become
zero.
Here we choose $\mu_S = \mu$ and calculate $\mu$ according to
Eq.~(\ref{mu_from_tau}) with $c_{\mathrm{max}}=1$, $n_L=1$, $n_H=3.3$ using
a $\nu_1$ coming from Eq.~(\ref{GegenbauerVisc}) with $\lambda=1$. The
$\tau$ in Eq.~(\ref{mu_from_tau}) is determined from $D$ alone using
Eq.~(\ref{min_tau}).

%
\begin{figure}
\centering
\includegraphics[width=0.99\linewidth]{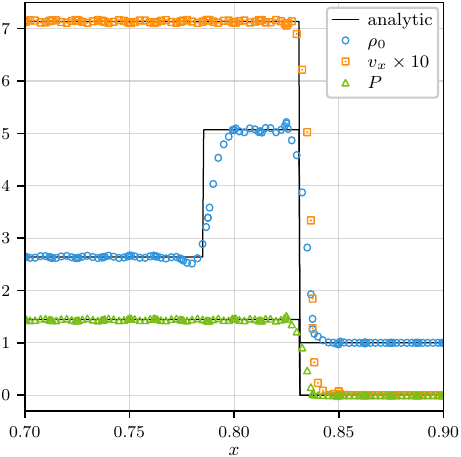}
\caption{
\label{riemannp_LN080t0.4_zoom}
Blast wave profiles for rest-mass density $\rho_0$, speed $v_x$,
and pressure $P$ at t = 0.4 after evolving the initial shocks in $\rho_0$
and $P$. The plot is for 80 elements covering $x\in[0,1]$
with 8 points per element.
We have zoomed in on the shock front, where the largest deviations from the
analytic solution (black solid lines) occur.
}
\end{figure}
The initial data has discontinuities in the matter fields $\rho_0$ and
$P$, as in a Riemann problem. We study a case from~\cite{Marti99},
that at $t=0$ is specified as
\be
\label{srhdID}
(\rho_0, v_x, P) = \left\{
\begin{array}{ll}
	(10.00, 0.00, 13.33)  & x\leq 0.5,  \\
	( 1.00, 0.00,  0.00)    & x > 0.5.
\end{array}
\right.
\ee
Here the density $\rho_0$, velocity $v_x$ and pressure $P$, as measured in
the fluid rest frame, comprise the primitive variables, from which we can
easily obtain the conservative variables $D$, $\tau$ and $S_j$.
This initial shock then propagates to the right. In
Fig.~\ref{riemannp_LN080t0.4_zoom} we show the shocked region at $t=0.4$.
We see that the fields in our simulation agree well with the analytic
solutions, plotted as solid black lines.
We should note that we have not carefully tuned the parameters
$c_{\mathrm{max}}$, $n_L$ and $n_H$, as we treat this test
more as a preliminary proof-of-concept vetting of our method.
For example, one could increase
$c_{\mathrm{max}}$ and obtain more viscosity. Alternatively, one could
increase $n_H$ so that the window in which some amount of viscosity
is applied gets wider. Both of these would suppress the oscillations,
that are still visible in the plot.
Nevertheless, we see that our AV method can successfully
handle evolutions of shock fronts in matter fields.

\subsubsection{Tests with stable neutron stars}

The next test starts from initial data for a static neutron star.
We then evolve this star in full GRHD with AV.
In these tests we have found that $S_j$ can become unstable inside the star
if we set $\mu_S = \mu$ and compute $\mu$ only from $D$, as in the shock
test above. We also find that the primitive variable $\epsilon$ is a more
sensitive sensor for discontinuities than $D$.
Therefore, we compute $\mu$ according to
Eq.~(\ref{mu_from_tau}) with $c_{\mathrm{max}}=0.5$, $n_L=1$, $n_H=3$,
using only $\epsilon$ to determine $\tau$ from Eq.~(\ref{min_tau}).
In order to set $\mu_S$, we first calculate a $\mu_{S_j}$ from
Eq.~(\ref{mu_from_tau}) with $c_{\mathrm{max}}=0.1$, $n_L=1$, $n_H=3$,
for each $S_j$, and finally set
$\mu_S = \max(\mu,\mu_{S_1},\mu_{S_2},\mu_{S_3})$.
For both $\mu$ and $\mu_S$ we choose a constant viscosity shape
($\nu(\bar{x})=1$) in each element.
We have observed that both the Gegenbauer and super Gaussian viscosities can
lead to instabilities, presumably because they drop to zero at element
boundaries, and thus do not provide sufficient dissipation there.

%
The particular star we consider is a Tolman-Oppenheimer-Volkoff
(TOV)~\cite{Tolman39,Oppenheimer39b} star with an ADM mass of
$m_{\mathrm{ADM}} = 1.655079$.
As equation of state (EoS) we use a piecewise
polytropic fit~\cite{Read:2008iy} of the zero-temperature MAP1
EoS~\cite{Muther:1987xaa} for which we add an additional ideal gas thermal
contribution~\cite{Shibata:2005ss} with
$\Gamma^{\rm hot}=1.75$~\cite{Bauswein:2010dn}.
With this setup we obtain a star with central density $\rho_{0,\mathrm{C}}=0.001215683$,
radius $r_*=8.295329$, and rest mass $m_0=1.872175$.

In order to evolve the gravitational fields we use the first order
generalized harmonic (GH) system~\cite{Lindblom:2005qh,Hilditch:2015aba},
with the gauge source functions $H_{\mu}$ set to zero. The constraint
damping parameters are chosen to be position dependent according to Eq.~(5)
of~\cite{Adhikari:2025nio}, where we use the modified
values $C_{(1)}=-1$, $A^{(1)}_{(3)}=0.999$, and $w^{(1)}_{(3)}=180$
for $\gamma_{1}$ .

We evolve this star on a mesh that consists of a central cube that is
surrounded by several shells made up of cube sphere patches. The outer
boundary is at a coordinate radius of $1096$.
We use the same constraint preserving outer boundary condition
as in~\cite{Lindblom:2005qh,Rinne:2007ui}.
To resolve the star well, we
use varying levels $l$ of h-refinement. The maximum amount of refinement
$l_{\mathrm{max}}$ occurs for points inside a sphere of radius
$r_{l_{\mathrm{max}}}=12$. As we move out, $l$ drops by one each time we
cross one of the radii $r_{l} = 12\times 2^{l_{\mathrm{max}}-l}$.

We employ the same artificial atmosphere treatment and primitive recovery
scheme as in~\cite{Adhikari:2025nio}, and also use the same positivity
limiters that are necessary for stable evolutions in low density regions.

\begin{figure}
\centering
\includegraphics[width=0.99\linewidth]{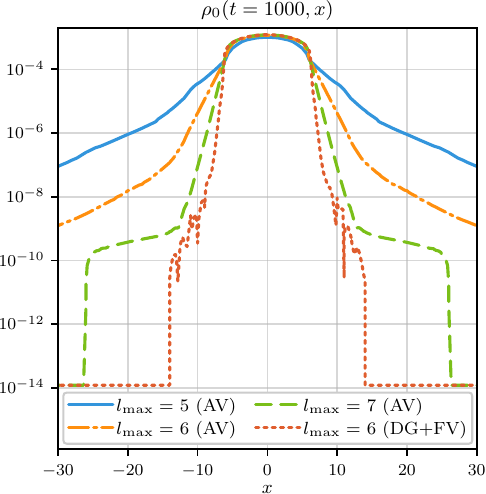}
\caption{
\label{rho0_vs_x}
The matter density $\rho_{0}$ at $t=1000$, for varying maximum
numbers $l_{\mathrm{max}}$ of h-refinement levels in the star. We
compare three different runs ($l_{\mathrm{max}}=5,6,7$) that use AV
with our standard hybrid DG+FV approach~\cite{Adhikari:2025nio}.
}
\end{figure}
In Fig.~\ref{rho0_vs_x} we show the density $\rho_{0}$ of the star after
we evolve until $t=1000$ for three different resolutions (solid,
dash-dotted, and dashed line), using our new AV approach. As we can see, the
star gets smeared out to some degree as one would expect with an AV method.
But the results improve quickly as we increase the maximum amount of
h-refinement $l_{\mathrm{max}}$. For comparison, we also show $\rho_{0}(x)$
when evolved with the hybrid scheme introduced in~\cite{Adhikari:2025nio},
that combines the DG method with a finite volume (FV) method.
To not overload the plot we show results only for $l_{\mathrm{max}}=6$
in the case of the hybrid DG+FV scheme (dotted line). As we can see,
the DG+FV scheme leads to a less smeared out star, because it uses
a FV method with shock capturing near the star surfaces. The amount of
h-refinement $l_{\mathrm{max}}=6$ for the hybrid scheme is lower than for
the highest $l_{\mathrm{max}}=7$ used in the AV approach. Note, however,
that the resolution in these two cases is comparable, because the hybrid
scheme uses more grid points in FV elements.

\begin{figure}
\centering
\includegraphics[width=0.99\linewidth]{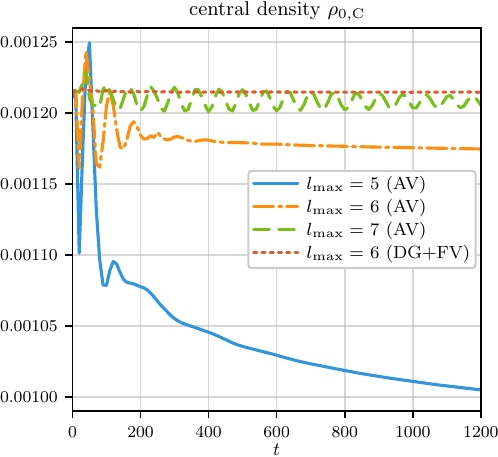}
\caption{
\label{rho0_AV_hybDG_FV}
The central density $\rho_{0,\mathrm{C}}$ versus time, for different runs
(see caption in Fig.~\ref{rho0_vs_x}).
}
\end{figure}
Fig.~\ref{rho0_AV_hybDG_FV} shows the central density $\rho_{0,\mathrm{C}}$ of the
star when we evolve with three different resolutions (solid, dash-dotted,
and dashed lines), using our new AV approach. We see that $\rho_{0,\mathrm{C}}$
approaches the expected constant behavior as we increase the resolution
$l_{\mathrm{max}}$. The hybrid DG+FV scheme is better in suppressing the
unphysical oscillations, that are visible with the AV approach, especially
near the beginning of the evolution.

\begin{figure}
\centering
\includegraphics[width=0.99\linewidth]{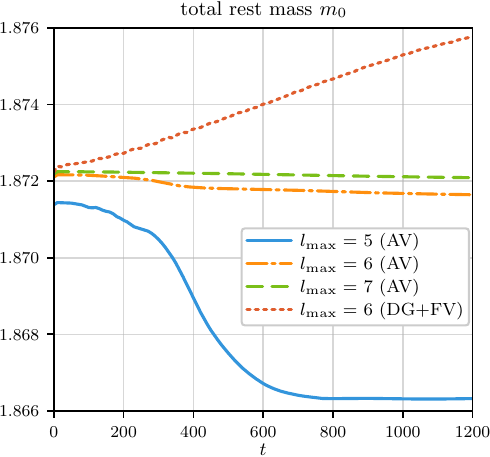}
\caption{
\label{restmass_AV_hybDG_FV}
The total rest mass versus time, for different runs
(see caption in Fig.~\ref{rho0_vs_x}).
}
\end{figure}
As we can see in Fig.~\ref{restmass_AV_hybDG_FV}, the rest mass conservation
improves with increasing resolution for the AV
approach, and is in fact better than for the hybrid DG+FV scheme.

\begin{figure}
\centering
\includegraphics[width=0.99\linewidth]{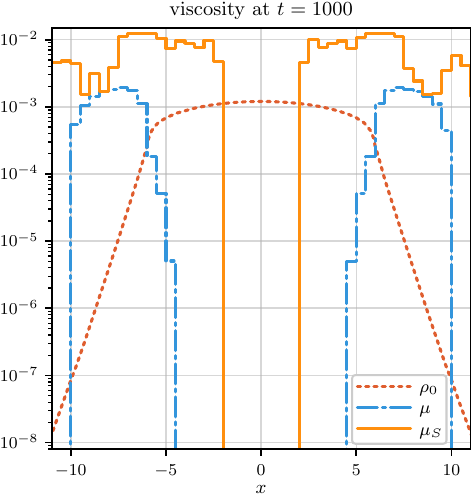}
\caption{
\label{viscosity_at_t1000}
The viscosities $\mu$, $\mu_S$ and the density $\rho_0$ after
evolving until $t=1000$.
}
\end{figure}
Fig.~\ref{viscosity_at_t1000} shows the viscosities $\mu$ (dash-dotted
line) and $\mu_S$ (solid line) along the $x$-axis at $t=1000$. To indicate
the star location we also show $\rho_0$ (dotted line). As we can see $\mu$
is non-zero only near the star surface where $D$ and $\rho_0$ vary sharply.
Notice, however that $\mu_S$ is non-zero in a larger region that reaches
into the higher density regions of the star.


\section{Discussion}
\label{discussion}

The DG formulation commonly used for hyperbolic evolution systems in
numerical-relativity codes such as
SpECTRE~\cite{Kidder:2016hev,Deppe:2021bhi} and
\nmesh~\cite{Tichy:2022hpa,Adhikari:2025nio} is formulated for systems that
are first order in space, i.e.~contain only first spatial derivatives.
In this paper, we describe in detail how an
LDG method can be formulated that can be used to evolve equations that
contain second spatial derivatives. This allows us to evolve equations like
the heat equation, which is genuinely second order in space, and
also equations like the wave equation in a from where we keep second order
derivatives. We have tested this LDG method with the scalar field equation
with second spatial derivatives, and also with the heat equation. We find
stable and convergent evolutions.

In this paper, we also describe in detail how this LDG method can be used to
add second order in space AV terms to first order in space hyperbolic
equations. This AV approach can then be used to treat systems with shocks and
other discontinuities. Our AV approach uses non-zero viscosity $\mu$ only in
elements where our new smoothness indicator $\tau$ (in Eq.~(\ref{min_tau}))
falls below a chosen value. In this way, no viscosity at all is present in
smooth regions, so that we achieve the full accuracy of a standard DG method
in smooth regions. We find that for neutron star evolutions it is important
to introduce two different viscosities $\mu$ and $\mu_S$ in
Eq.~(\ref{GRHD_with_av}). These two viscosities are computed by considering the
smoothness of the primitive variable $\epsilon$ and the conserved $S_i$
respectively.

We show successful evolutions with AV for test cases such as Burgers'
equation, special relativistic blast wave tests, and the evolution of single
TOV stars in full GRHD.
Our new AV method works in all cases considered. It
has several advantages, namely, it is easy to implement, is simpler than
hybrid FV/DG, and it is faster than hybrid FV/DG, because elements with FV
use more points in our hybrid FV/DG method. On the other hand, LDG requires
more communication. Also, as we have seen in Fig.~\ref{viscosity_at_t1000},
$\mu_S$ is non-zero in a region that reaches into the higher density regions
of the star. This behavior may not be desired, as it might dissipate too much
momentum in case the star is moving. Further research is needed to determine
if this is indeed a problem, and whether we can obtain a more localized
$\mu_S$ by adjusting parameters such $c_{\mathrm{max}}$, $n_L$ and
$n_H$.

To the best of our knowledge this is the first time that a neutron star in
full GRHD has been successfully evolved in full 3D with a DG method using AV.
Prior works such as~\cite{Anninos_2017} have not attempted to simulate
self-gravitating systems such as neutron stars. Similarly, the recent work
in~\cite{Atteneder2025a} made substantial progress for simpler test cases,
but was not able to stably evolve neutron stars.

%
%

\acknowledgments

It is a pleasure to thank Florian Atteneder for helpful discussions.
This work was supported by NSF grants
PHY-2136036 and PHY-2408903.
It was also supported by PeX-FCT (Portugal) Grant No. 2023.12549.PEX.
We also acknowledge usage of computer time on the HPC cluster Athene at
Florida Atlantic University.



%
%
%
% ...
%
%
% ...
%
%
%
%
%



\bigskip
\noindent

\bibliography{references}

\end{document}